\documentstyle[12pt]{article}
\newcommand{\ket}[1]{\left | #1 \right \rangle}

\begin{document}

\noindent
{\small Appearing in {\em Geometric Issues in the Foundations of
Science} eds. S. Huggett, L. Mason, K. P. Tod, S. T. Tsou and
N. M. J. Woodhouse, Oxford University Press 1997.} 
\begin{center}
{\Large\bf Entanglement and Quantum Computation} \bigskip \\
Richard Jozsa\\ School of Mathematics and
Statistics, \\University of Plymouth, Plymouth, Devon PL4 8AA, England.
\\Email: rjozsa@plymouth.ac.uk
\end{center}
\bigskip
\noindent
{\large\bf Introduction}\\[3mm]
The phenomenon of quantum entanglement is perhaps the most enigmatic
feature of the formalism of quantum theory. It underlies many of the
most curious and controversial aspects of the quantum mechanical description 
of the world. In \cite{PEN} Penrose gives a delightful and accessible 
account of entanglement illustrated by some of its extraordinary
manifestations. Many of the best known features depend on issues of
{\em non-locality}. These include the seminal work of Einstein, Podolsky
and Rosen \cite{EPR}, Bell's work \cite{BELL} on the EPR singlet state,
properties of the GHZ state \cite{GHZ,MER} and Penrose's dodecahedra
\cite{PEN}. In this paper we describe a new feature of entanglement
which is entirely independent of the auxiliary
notion of non-locality. 

We will argue that the phenomenon of entanglement is responsible for
an essential difference in the {\em complexity} (as quantified below)
of physical evolution allowed by the laws of quantum physics
in contrast to that allowed by
the laws of classical physics. This distinction was perhaps 
first explicitly realised by Feynman \cite{FEY} in 1982 when he noted that
the simulation of a quantum evolution on a classical computer
appears to involve an unavoidable exponential slowdown in running time.
Subsequently in the development of the subject of quantum computation --
which represents a hybrid of quantum physics and theoretical computer
science -- it was realised that quantum systems could be harnessed to perform
useful computations more efficiently than any classical device.
Below we will examine some of the basic ingredients of quantum
computations and relate their singular efficacy to the existence of
entanglement.

The perspective of information theory also provides further new
insights into the relationship between entanglement and non-locality,
beyond the well studied mediation of non-local correlations between
local measurement outcomes. To outline some of these effects we first 
introduce the notion of ``quantum information''.

A fundamental difference between quantum and classical physics is that
the state of an unknown quantum system is in principle unmeasureable
e.g. given an unknown state $\ket{\psi}$ of a 2-level system it is
not possible to identify it. In fact any measurement on $\ket{\psi}$ 
will reveal at most one bit of information about its identity whereas
the full description of $\ket{\psi}$ requires the specification of
two real numbers. In terms of binary expansions this corresponds to a 
double infinity of bits of information. We refer to the full (largely
inaccessible) information represented by a quantum state as quantum 
information in contrast to the more familiar notion of classical information
such as the outcome of a measurement which is in principle fully
accessible.

The inaccessibility of quantum information is closely related to
the possibility of non-local influences which do not violate
classical causality, as necessitated by Bell's \cite{BELL} analysis
of local measurements on an EPR pair -- the non-local influences
are simply restricted to a level which is inaccessible to any
local observations. However it has recently been shown \cite{TELE,CONC}
that entanglement plays a more subtle role here than just mediating
correlations between the classical information of local measurement outcomes.
According to the process known as quantum teleportation \cite{TELE}
entanglement acts as a {\em channel} for the transmision of quantum
information: an unknown quantum state of a 2-level system may be
transferred intact from one location to another by sending only two bits
of classical information, if the locations are also linked by
entanglement in the form of a shared EPR pair. The remaining quantum
information can be interpreted as having flown across the entanglement
which is destroyed by the process.

Another novel application of entanglement and non-locality
inspired by concepts from theoretical computer science, 
is the existence and construction of quantum error correcting codes,
first introduced by Shor \cite{ERR} in 1995.
Entanglement provides a way of delocalising quantum information in
a system composed of several subsytems. For example if $\ket{0}$ and
$\ket{1}$ are orthogonal states then the entangled states
$\frac{1}{\sqrt{2}}( \ket{0}\ket{0}+ \ket{1}\ket{1})$ and 
$\frac{1}{\sqrt{2}}( \ket{0}\ket{1}+\ket{1}\ket{0})$
are identical in terms of local quantum information (each subsystem being
in the maximally mixed state in each case) whereas they differ in terms of 
their global quantum information content. By an ingenious extension of
this idea \cite{ERR} one may encode an unknown state of a 2-level system
into an entangled state of several 2-level systems in such a way that if 
any (unknown) one of the subsystems is arbitrarily corrupted then the 
original state may still be perfectly reconstructed i.e. none of the 
information of the original state resides locally in the encoding.
In this way a state
may be protected against the effects of spurious environmental interactions
if (as is generally a good approximation) these interactions act by local
means.     

Thus quantum computation and quantum information theory provide a rich variety
of new applications of entanglement and we now turn to a more detailed
discussion of issues in quantum computation in particular.  
\\[3mm]
\noindent
{\large\bf Quantum Computation and Complexity}\\[3mm]
The theory of computation and computational complexity \cite{PAP}
is normally presented as an entirely mathematical theory with no
reference to considerations of physics. However any actual
computation is a physical process involving the physical evolution of 
selected properties of a physical system. Consequently the issues of
``what is computable'' and ``what is the complexity of a computation''
must depend essentially on the laws of physics and cannot be characterised
by mathematics alone. This fundamental point was emphasised by
Deutsch \cite{DD1} and it is dramatically confirmed by the recent
discoveries \cite{DJ,BV,SI,SH,GR,KIT}
that the formalism of quantum physics allows one to transgress
some of the boundaries of the classical theory of computational
complexity, whose formulation was based on classical intuitions.  
In \cite{PEN} Penrose proposes the possible introduction of
non-computable elements into physics (i.e. non-computable within
the standard existing theory of computability). This however requires 
going beyond the existing formalism of quantum theory since the latter
lies entirely within the bounds of classical computability. Our considerations
here lie entirely within the standard framework so that, for us, 
quantum processes 
cannot result in any computation which is not already possible by classical
means. This notwithstanding, there does appear to be a 
significant difference in the
{\em efficiency} of computation as noted in Feynman's remark.

A fundamental notion of the theory of computational
complexity is the distinction between polynomial and exponential use of
resources in a computation. 
This will provide a quantitative measure of our essential distinction
between quantum and classical computation.
Consider a computational task such as the following: given an integer $N$,
decide whether $N$ is a prime number or not. We wish to assess the resources 
required for this task as a function of the size of the input which is
measured by $n=\log_2 N$, the number of bits needed to store $N$.
If $T(n)$ denotes the number of steps (on a standard universal computer)
needed to solve the problem, we ask whether $T(n)$ can be bounded
by some polynomial function in $n$ or whether $T(n)$ grows faster
than any polynomial. More generally we may consider any language
$\cal L$ -- a language being a subset of the set of all finite
strings of 0's and 1's -- and consider the computational task of recognising
the language i.e. given a string $\sigma$ of length $n$ the computation 
outputs 0 if $\sigma \in {\cal L}$ and outputs 1 if $\sigma \not\in 
{\cal L}$. (In our example above $\cal L$ is the set of all prime numbers 
written in binary). The language $\cal L$ is said to be in complexity class
$\cal P$ (``polynomial time'') if there exists an algorithm which
recognises $\cal L$ and runs in time $T(n)$ bounded by a polynomial 
function. Otherwise the recognition of $\cal L$ is said to require
exponential time. More generally it is useful to consider algorithms
which involve probabilistic choices (``coin tosses'') \cite{PAP,EJ}.
$\cal L$ is said to be in the class $\cal BPP$ (``bounded-error
probabilistic polynomial time'') if there is a polynomial time algorithm 
which correctly classifies the input string $\sigma$ with probability
at least 2/3 (or equivalently, any other value strictly between 1/2 and 1).
Thus a $\cal BPP$ algorithm may give the wrong answer but by simple repetition
and taking the majority answer, we can amplify the probability of success as 
close to 1 as desired while retaining a polynomial time for the whole
process (\cite{PAP,EJ}).
The class $\cal BPP$ of algorithms which run in polynomial time but allow 
for ``small imperfections'', is often regarded as the class of computational
tasks which are ``feasible in practice'' (or at least a first
mathematical approximation to this notion). We also use the term
``efficient computation'' for a computation which runs in polynomial time.

In the above considerations the exact number of steps $T(n)$   
will generally depend on the choice of underlying computer and the
model of computation adopted. However if we stay within the confines of
classical physics, the distinction between polynomial and
exponential time i.e. between efficient and non-efficient computation,
appears to be robust, being independent of these choices.
It is thus a distinction with {\em physical} significance.
In the discussion above we have illustrated it in its most
familiar form -- in terms of the resource of time. From the physical
point of view it is natural to extend the notion of efficient computation
to require the efficient use of {\em all} possible physical resources.  
Indeed in the following discussion we will be led to consider other
resources such as energy. The absolute significance of the distinction
between efficient and non-efficient computation provides an extension
of the classical Church-Turing thesis \cite{DD1,SH} which refers
to a similar distinction between computability and non-computability.
The fundamental {\em raison d'\^{e}tre} of quantum computation is
the fact that quantum physics appears to allow one to transgress
this classical boundary between polynomial and exponential
computations. 

The concept of 
quantum computation may be rigorously formalised as a natural extension
of classical mathematical models of computation \cite{DD1,BV,SH,YAO,EJ}
in which the computational steps are allowed to be quantum processes
restricted by a suitable notion of locality. For our purposes
it will suffice to envisage a quantum computer as a standard universal
computer in which the memory bits are 2--level quantum systems
instead of 2--state classical systems. The quantum systems are each endowed
with a preferred basis $\{ \ket{0}, \ket{1} \}$ corresponding to the
classical bit values of 0 and 1. We refer to these 2--level systems
as qubits \cite{SCHUM}.

The computer is able to support arbitrary superpositions of the values 0 and 
1 within each qubit and also entanglements of many qubits.
Furthermore the computer may be programmed to perform unitary transformations 
of any number of qubits. It is important however that large unitary 
transformations be constructed or ``programmed'' from a finite set of fixed 
given unitary transformations. In this way we can assess the complexity
of unitary transformations by the length of their programs.
There are many known examples of small finite sets of transformations, 
out of which one can program any unitary transformation of any
number of qubits (to arbitrary precision) \cite{DD2,NINE,BAR}.
Indeed it is known \cite{BDE} that almost any single transformation of two
qubits by itself suffices. The distinction between polynomial 
and exponential time does not depend on the choice of these basic
transformations as any one such set can first be used to build the 
members of any other set leading to only an overall constant expansion in 
the number of elementary steps in any computation. \\[3mm]
{\large\bf Superposition and Entanglement in Quantum Computation}\\[3mm]
There are several quantum algorithms known \cite{DJ,BV,SI,SH} which 
strongly support the view that a quantum computer can perform
some computational tasks exponentially faster than any classical
device. The most significant of these is Shor's polynomial time algorithm
for integer factorisation \cite{SH,EJ}, a problem which is believed to
lie outside the class $\cal BPP$ of classical computation. We then ask:
what is the essential quantum effect that gives rise to this exponential 
increase in computing power?

All of the quantum algorithms utilise the process of computation
by quantum parallelism \cite{DD1}. This refers to a quantum computer's 
capability to carry out many computations simultaneously in superposition
if the input is set up in a suitable superposition of classically  
distinct inputs. Thus one might conclude that it is {\em superposition} 
that is at the root of the quantum computational speedup. However closer
examination will show that {\em entanglement} is the essential feature
rather than just superposition itself. Note that entanglement may be viewed as
a special kind of superposition -- superposition in the presence of a 
product structure on the state space -- which arises from the system
being made up of several subsystems. In our considerations these are the qubits
comprising the computer. An entangled state is then a superposition
of product states which cannot be expressed as a single product state.

To see that superposition itself is not the essential feature we need
only note that classical waves also exhibit superposition. Any effect
depending on quantum interference alone can be readily mimicked by
classical waves. However in other respects quantum states and classical waves
differ considerably e.g. the measurement theories are very different
(being far more favourable for computation with classical waves than
with quantum states!) and {\em there is no classical analogue of
the phenomenon of entanglement}.

To illustrate the above remarks and highlight the role of entanglement
consider the following example of computation by quantum parallelism.
Let $B =\{ 0,1 \}$ and consider any (non-trivial) function
$f:B^n \rightarrow B$. Suppose that we have a quantum computer programmed
to compute $f$ in polynomial time. The computer has $n$ input qubits and
one output qubit and its operation corresponds to a unitary transformation
${\cal U}_f$ on $n+1$ qubits which effects the evolution:
\[ {\cal U}_f:\,\,
\underbrace{\ket{i_1}\ket{i_2} \ldots \ket{i_n}}_{\rm input}
\ket{0} \longrightarrow
   \ket{i_1}\ket{i_2} \ldots \ket{i_n}\ket{f(i_1 ,\ldots ,i_n )} \]
Here each $i_k$ is either 0 or 1. The output register is initially in
state $\ket{0}$ and at the end of the computation it contains the basis state 
corresponding to the value of the function.
Consider the one-qubit operation:
\[ H= \frac{1}{\sqrt{2}} \left( 
\begin{array}{cc} 1 & 1 \\
          1 & -1
\end{array} \right)  \]
If the input qubits are all initially in state $\ket{0}$ then
applying $H$ to each one successively gives an equal superposition of
all $2^n$ values in $B^n$:
\begin{equation} \label{one}
H\otimes \cdots \otimes H \,\,\ket{0}\cdots \ket{0} = \frac{1}{2^{n/2}}
\left( \ket{0}+\ket{1} \right)^n =  \frac{1}{2^{n/2}}
\sum_{i=0}^{2^n -1} \ket{i} 
\end{equation}
(where we have identified $(i_1 , \ldots ,i_n )$ with the binary
number $i_1 \ldots i_n < 2^n$).
Note that this state is prepared in polynomial (in fact linear) time.
Running the computer with (\ref{one}) as input yields the final state
\begin{equation} \label{two}
\ket{f} = \sum_{i=0}^{2^n -1} \ket{i}\ket{f(i)}
\end{equation}
Thus by quantum parallelism we have computed exponentially many values of
$f$ in superposition with only polynomial computing effort.

Can we mimick the above with classical waves? We represent each qubit by
a classical wave system and select two modes of vibration to represent
the states $\ket{0}$ and $\ket{1}$ e.g. $\ket{0}$ and $\ket{1}$ may
be the two lowest energy modes of a vibrating elastic string with
fixed endpoints. It is then straightforward to construct the superposition
corresponding to $\ket{0}+\ket{1}$ and perorming this separately on
$n$ pieces of string we obtain the product state (\ref{one}).
However, regardless of how much the strings interact with each other
in their subsequent (externally driven) vibrational evolution, their
joint state is always a {\em product} state of $n$ separate
vibrations. The total state space of the total classical system is
the {\em Cartesian} product of the individual state spaces of the
subsystems whereas quantum-mechanically, it is the {\em tensor}
product. This crucial distinction between Cartesian and tensor products
is precisely the phenomenon of quantum entanglement.
The state (\ref{two}) is generally entangled (for non-trivial $f$'s)
so that the transition from (\ref{one}) to (\ref{two}) cannot be
achieved in a classical scenario.

We may attempt to represent entanglement using classical waves   
in the following manner. The state of $n$ qubits is a $2^n$ dimensional space 
and can be isomorphically viewed as the state space of a {\em single}
particle with $2^n$ levels. Thus we simply interpret certain states
of a single $2^n$ level particle as ``entangled'' via their correspondence
under a chosen isomorphism between $\bigotimes^n {\cal H}_2 $ and
${\cal H}_{2^n}$ (where ${\cal H}_k$ denotes a Hilbert space of dimension
$k$.) In this way, $2^n$ modes of a classical vibrating system can
apparently be used mimick general entanglements of $n$ qubits. 
However the physical implementation of this correspondence appears always to
involve an exponential overhead in some physical resource so that the 
isomorphism is {\em not} a valid correspondence for considerations of
complexity. For example suppose that the $2^n$ levels of the 
one-particle quantum system 
are equally spaced energy levels.  
A general state of $n$ qubits requires an amount of energy that grows 
{\em linearly} with $n$ whereas a general state of this $2^n$ level system 
(and also the corresponding classical wave system)
requires an amount of energy that grows {\em exponentially} with $n$.
To physically realise a system in a general  superposition of $2^n$ modes
we need exponential resources classically and linear resources
quantum mechanically {\em because of the existence of entanglement}.

This comparison is reminiscent of the representation of
whole numbers in unary (i.e. representing $k$ as a string of $k$ 1's
analogous to $k$ equally spaced levels) versus the binary representation
of $k$ which requires a string of length $\log_2 k$ and is therefore
exponentially more efficient.
Note that $n$ classical bits can also accommodate $2^n$ possible alternatives
but only one such alternative can be present at any time, even if
probabilistically determined. In contrast $n$ qubits can accommodate
$2^n$ possible alternatives which may all be {\em simultaneously}
present in superposition. As a quantum computation proceeds a new qubit may
be incorporated into the overall processed state at each step leading to
an exponential growth in time of the quantum information,
because of entanglement. If we were to compute this quantum evolution 
``by hand'' from the laws of quantum mechanics then we would suffer an 
exponential slowdown in handling an exponentially growing amount of 
information. In contrast, Nature manages to process this growing
information in linear time! This is an example of Feynman's remark 
mentioned in the introduction. Note that if the state of the accumulating
qubits were always a product state (i.e. no entanglement) 
then the quantum information would accumulate only linearly.

There exist physical systems with infinitely many discrete energy levels
which accumulate below a finite bound $E_0$.
Thus we may use these levels to represent general superpositions of
exponentially many modes with only a constant cost in energy
and apparently circumvent the above objections!
However in this case the levels will crowd together exponentially closely
and we will need to build our instruments with exponentially finer precision.
This again will presumably require exponentially increasing physical
resources.

It has been occasionally suggested that the interferences  
in Shor's efficient quantum factoring algorithm and other quantum
algorithms, can be readily represented by classical wave interferences
but on closer inspection all such proposals involve an exponential overhead
in some physical resource as illustrated in our discussion above.

The standard mathematical theory of computational complexity \cite{PAP}
assesses the complexity of a computation in terms of the resources of
time (number of steps needed) and space (amount of memory required).
In the above we have been led to consider the accounting of other physical
resources such as energy and precision\cite{SH}. This reinforces our earlier 
remark that the notion of computational complexity must rest  
on the laws of physics and consequently the proper assessment of complexity
will need to take into account all possible varieties of physical
resource. A theory of computational complexity based on such general 
physical foundations remains to be formulated.
\\[3mm]
{\large\bf Entanglement and the Super-Fast Quantum Fourier Transform}\\[3mm]
The efficiency of Shor's factoring algorithm rests largely on
the fact that the discrete Fourier transform\cite{EJ} $DFT_{2^n}$ 
in dimension $2^n$ (a particular unitary transformation in $2^n$
dimensions) may be implemented on a quantum computer in time
polynomial in $n$ (in fact quadratic in $n$). Similarly the efficiency of
Deutsch's and Simon's algorithms \cite{DJ,SI}
rest on the polynomial-time computability
of the Fourier transform over the additive group $B^n$.
The standard classical Fourier transform algorithm implements
$DFT_{2^n}$ in time $O((2^n)^2)$ and the classical {\em fast} Fourier 
transform algorithm improves this to $O(n2^n )$ which is an exponential 
saving but still remains exponential in $n$. We will argue that
the extra quantum mechanical speedup to $O(n^2 )$ resulting in a
genuinely polynomial-time algorithm, is due to the effects of
entanglement. We will describe a simplified example which illustrates the
essential principle involved.

Let $M$ be a unitary matrix of size $2^n$ and $v$ a column vector
of length $2^n$. To compute $w=Mv$ by direct matrix multiplication
requires $O((2^n)^2)$ operations, each entry of the result requiring
$O(2^n)$ operations. 
Suppose now that the space of $v$ is the tensor product of $n$
two dimensional spaces $V^{(1)}\otimes \ldots \otimes V^{(n)}$ and that $M$ 
decomposes as a simple tensor product
\begin{equation} \label{tp}
M = S^{(1)} \otimes \ldots \otimes S^{(n)} \end{equation}
where each $S^{(i)}$ is a 2 by 2 matrix acting on the respective
component space $V^{(i)}$. Thus we can label the components of $v$ by 
indices $i_1 \ldots i_n \in B^n$ and the computation of $w$ becomes
\begin{equation} \label{prod}
w_{j_1 \ldots j_n} = \sum_{i_1 \ldots i_n} 
S^{(1)}_{j_1 i_1} \ldots S^{(n)}_{j_n i_n} \, v_{i_1 \ldots i_n}
\end{equation}
Now consider first $S^{(1)}$. Each application of this 2 by 2 matrix requires 
some fixed (i.e. independent of $n$) number of operations
and $S^{(1)}$ needs to be applied $2^{n-1}$ times, once for each
choice of the indices $i_2 \ldots i_n \in B^{n-1}$. The same accounting 
applies to each of the $n$ matrices $S^{(i)}$ giving a
total number of operations $O(n2^{n-1}) = O(n2^n )$. 
Thus the tensor product factorisation (\ref{tp}) leads to an exponential 
speed-up compared to straightforward matrix multiplication for a 
general $M$. A similar argument will apply if $M$ decomposes
more generally into the successive application of a polynomial number
of matrices, each of which applies to a bounded number $b$ of the
component spaces (not necessarily disjoint) and $b$ is independent of $n$. 
(\ref{tp}) is merely the simplest example of such a decomposition. 
The classical fast Fourier transform algorithm is based on the fact
that the Fourier transform in dimension $2^n$ decomposes in just this way
(although not as simply as (\ref{tp}) c.f. \cite{EJ} for an explicit
description of the decomposition).

Consider finally the implementation of $M$ as given in (\ref{tp}) 
on a quantum computer. The data given by the components of $v$ is
represented by the amplitudes of a general state of $n$ qubits.
Each of the $n$ operations $S^{(i)}$ is a one-qubit operation
and needs to be applied {\em only once} to its respective qubit
i.e. the $2^{n-1}$ repetition of the classical calculation
is eliminated! This is due to the rules of formation of the tensor
product (i.e. entanglement) requiring for example that $S^{(1)}$
applied to the first qubit automatically carries through
to all possible values of the indices $i_2 \ldots i_n$
in (\ref{prod}) i.e. the global operation $S^{(1)} \otimes I \otimes
\ldots \otimes I$ is implemented on the whole space.
 Thus $M$ is implemented in time $O(n)$.
In a similar way, the more complicated decomposition of
the Fourier transform can be seen \cite{EJ} to lead to a time $  
O(n^2 )$.
This comparison of classical and quantum implementations of $M$
is yet another illustration 
of Feynman's remark \cite{FEY} that the simulation of a quantum
process on a classical computer generally involves an exponential slowdown.

The classical and quantum scenarios for the above computation of
$Mv$ differ significantly in the following respect.
After the classical computation we are able to read off all $2^n$
components of $w$ presented as classical information
whereas the quantum computation results in the quantum information of
one copy of the state $\ket{w}= \sum w_{i_1 \ldots i_n} \ket{i_1}
\ldots \ket{i_n}$ from which we are unable to extract the individual
values of $w_{i_i \ldots i_n}$. 
This is the issue of inaccessibility of quantum information that
was mentioned in the introduction.
Nevertheless we are able to extract classical information 
from $\ket{w}$ that depends on 
exponentially many of the values and classically this information would
require a preliminary exponential computational effort.
This phenomenon also occurs in computation by quantum parallelism.
From the quantum information $\ket{f}$ in (\ref{two}) we are unable
to extract all the individual values $f(i_1 \ldots i_n )$ but we
can obtain certain global properties of the function. Indeed in Shor's
factoring algorithm \cite{SH,EJ}, the analogue of $f$ 
is a periodic function and after applying the Fourier
transform we are able to extract the period.\\[2mm]
{\large\bf Concluding Remarks}\\[2mm]
In summary the effects of quantum entanglement enable certain large
unitary transformations to be implemented exponentially more efficiently
on a quantum computer than on any classical computer. However after quantum
computation the full results are coded in a largely inaccessible form.
Remarkably this inaccessibility, dictated by the principles of
quantum measurement theory, does {\em not} serve to cancel out the
exponential gain in computing effort.  
Limited information may be obtained
about the transformed data which, although small, 
would nevertheless require an exponential
amount of computing effort to obtain by classical computation. 

Another fundamental issue intimately related to entanglement is the
so-called measurement problem of quantum mechanics. This refers to 
the reconciliation of the apparent ``collapse of wave function''
in a measurement with the unitary evolution of quantum mechanics and
explaining why, after a measurement, we see merely one of the possible 
outcomes rather than experiencing some weird reality in entanglement
with all possible outcomes. In \cite{PEN} Penrose discusses several of the 
best known interpretations of quantum mechanics and it appears that none 
of them provides a resolution of this phenomenon.
With extraordinarily innovative and broadranging arguments, Penrose 
suggests that the resolution might involve non-computational
ingredients. The fact that quantum theory has resisted unification with
the theory of gravitation suggests that the essentially linear
concept of entanglement may not persist at a macroscopic level.
Indeed it is difficult to imagine that the linearity of quantum theory
could survive a unification with the nonlinear foundations of the
general theory of relativity.
The algorithms of quantum computation such as Shor's algorithm
depend critically for their efficiency and validity on effects
of increasingly large scale entanglements with increasing input
size. Thus efforts to experimentally implement these
algorithms may provide particularly acute opportunities to witness
a possible breakdown of the current conventional quantum formalism.


\begin{thebibliography}{xx} 
\bibitem{PEN} Penrose, R. (1994) {\em Shadows of the Mind} (Oxford
University Press). 
\bibitem{DJ}  Deutsch, D. and Jozsa, R. (1992) {\em Proc. Roy. Soc. London
Ser A} {\bf 439}, 553-558.
\bibitem{BV} Bernstein, E. and Vazirani, U. (1993)
{\em Proc. 25th Annual ACM Symposium on the Theory
of Computing}, (ACM Press, New York), p. 11-20
(Extended Abstract). Full version of this paper appears in
{\em S. I. A. M. Journal on Computing} {\bf 26} (Oct 1997). 
\bibitem{SI} Simon, D. (1994) {\em Proc. of 35th Annual Symposium on the
Foundations of Computer Science}, (IEEE Computer Society, Los Alamitos),
p. 116 (Extended Abstract). Full version of this paper appears in
{\em S. I. A. M. Journal on Computing} {\bf 26} (Oct 1997).
\bibitem{SH} Shor, P. (1994) {\em Proc. of 35th Annual Symposium on the
Foundations of Computer Science}, (IEEE Computer Society, Los Alamitos),
p. 124 (Extended Abstract). Full version of this paper appears in
{\em S. I. A. M. Journal on Computing} {\bf 26} (Oct 1997) and is also
available at LANL quant-ph preprint archive 9508027.
\bibitem{GR} Grover, L. (1996) 
{\em Proc. 28th Annual ACM Symposium on the Theory
of Computing}, (ACM Press, New York), p. 212-219.
\bibitem{PAP} Papadimitriou, C. H. (1994) {\em Computational Complexity}
(Addison-Wesley, Reading, MA). 
\bibitem{DD1} Deutsch, D. (1985) {\em Proc. Roy. Soc. London Ser. A} 
{\bf 400}, 97. 
\bibitem{EJ} Ekert, A. and  Jozsa, R. (1996) {\em Rev. Mod. Phys.} {\bf 68},
733.
\bibitem{KIT} Kitaev, A. (1995) {\em ``Quantum Measurements and 
the Abelian Stabiliser
Problem''}, preprint available at LANL quant-ph preprint archive 9511026.
\bibitem{YAO} Yao, A. (1993) {\em Proceedings of the 34th Annual Symposium
on the Foundations of Computer Science}, edited by S. Goldwasser,
IEEE Computer Society, Los Alamitos, CA. p. 352.
\bibitem{SCHUM} The term {\em qubit} was first introduced in
Schumacher, B. (1995) {\em Phys. Rev. A} {\bf 52}, 2738. 
\bibitem{DD2} Deutsch. D. (1989) {\em Proc. Roy. Soc. London Ser. A}
{\bf 425 }, 73.
\bibitem{NINE} Barenco, A., Bennett, C., Cleve, R., DiVincenzo, D., 
Margolus, N., Shor, P., Sleator, T., Smolin, J. and Weinfurter, H. (1995)
{\em Phys. Rev. A} {\bf 52}, 3457.   
\bibitem{BAR} Barenco, A. (1995) {\em Proc. Roy. Soc. London Ser. A} 
{\bf 449}, 679.
\bibitem{BDE} Barenco, A.,  Deutsch, D. and Ekert, A. (1995)
{\em Proc. Roy. Soc. London Ser. A} {\bf 449}, 669. 
\bibitem{FEY} Feynman, R. (1982) {\em Int. J. Theor. Phys.} {\bf 21}, 467.
\bibitem{EPR} Einstein, A., Podolsky, B. and Rosen, N. (1935)
{\em Phys. Rev.} {\bf 47}, 777.
\bibitem{BELL} Bell, J. S. (1964) {\em Physics} {\bf 1}, 195. 
\bibitem{GHZ} Greenberger, D., Horne, M. and Zeilinger, A. (1989)
in {\em Bell's theorem, quantum theory and conceptions of the universe}
(ed. M. Kafatos) p73 (Kluwer Academic, Dordrecht, The Netherlands). 
\bibitem{MER} Mermin, N. D. (1990) {\em Am. J. Phys.} {\bf 58}, 731.
\bibitem{TELE} Bennett, C., Brassard, G., Crepeau, C., Jozsa, R.,
Peres, A. and Wootters, W. (1993) {\em Phys. Rev. Lett.} {\bf 70}, 1895.
\bibitem{CONC} Bennett, C., DiVincenzo, D., Smolin, J. and Wootters, W.
(1996) {\em Phys. Rev. A} {\bf 54}, 3824.
\bibitem{ERR} Shor, P. (1995) {\em Phys. Rev. A} {\bf 52}, R2493.

\end{thebibliography}
\end{document}